\documentclass[11pt]{article}

\usepackage{amssymb}
\usepackage{amsmath}
\usepackage{graphicx}
\usepackage{latexsym}
\usepackage{subfigure}

\begin{document}

\begin{titlepage}
\begin{center}

\hfill \today

\vspace{0.5cm}
{\large\bf 130GeV gamma-ray line through axion conversion}
\vspace{1cm}

{\bf Masato Yamanaka}$^{a, b\,}
$\footnote{yamanaka@eken.phys.nagoya-u.ac.jp}, 
{\bf Kazunori Kohri}$^{b, c,\,}
$\footnote{kohri@post.kek.jp},
{\bf Kunihito Ioka}$^{b, c,\,}
$\footnote{kunihito.ioka@kek.jp},

{\bf and Mihoko M. Nojiri}$^{b, c,\,}
$\footnote{nojiri@post.kek.jp} 

\vskip 0.15in

{\it
$^a${Department of Physics, Nagoya University, Nagoya 464-8602, Japan}\\
$^b${Theory Center, Institute of Particle and Nuclear Studies,
KEK (High Energy Accelerator Research Organization),
1-1 Oho, Tsukuba 305-0801, Japan}\\
$^c${The Graduate University for Advanced Studies (Sokendai), 
1-1 Oho, Tsukuba 305-0801, Japan}
}

\vskip 0.4in

\abstract{  We apply the axion-photon conversion mechanism to the 130 GeV 
$\gamma$-ray line observed by the Fermi satellite. Near the Galactic center, 
some astrophysical sources and/or particle dark matter can produce
energetic axions (or axionlike particles), and the axions convert to $\gamma$ 
rays in Galactic magnetic fields along their flight to the Earth. Since continuum 
$\gamma$-ray and antiproton productions are sufficiently suppressed in 
axion production, the scenario fits the 130 GeV $\gamma$-ray line without 
conflicting with cosmic ray measurements. We derive the axion production 
cross section and the decay rate of dark matter to fit the $\gamma$-ray 
excess as functions of axion parameters. In the scenario, the $\gamma$-ray 
spatial distributions depend on both the dark matter profile and the magnetic 
field configuration, which will be tested by future $\gamma$-ray observations, 
e.g., H.E.S.S. II, CTA, and GAMMA-400. As an illustrative example, we study 
realistic supersymmetric axion models, and show the favored parameters 
that nicely fit the $\gamma$-ray excess.}

\end{center}
\end{titlepage}

\tableofcontents

\section{Introduction} \label{Sec:Intro}   

Cosmological and astrophysical measurements have been supporting 
the existence of dark matter~\cite{Clowe:2006eq, Hinshaw:2012fq}. 
Numerical simulations~\cite{Navarro:1995iw, Springel:2008cc, 
BoylanKolchin:2009nc, Guo:2010ap} and the thermal relic 
scenario~\cite{Lee:1977ua, Kolb} suggests that weakly interacting 
massive particles (WIMPs) are the most likely candidate for dark matter.
Indirect detection, which searches for the annihilation and/or decay 
products of dark matter, is a promising way to explore its nature. 
Indirect detection is quite important in checking thermal relic 
scenarios of WIMP dark matter, because the annihilation cross 
section directly connects with its thermal relic abundance.

One of the aims of the Fermi telescope is to search for $\gamma$-ray 
signals from dark matter~\cite{Baltz:2008wd}.
An excess of $\gamma$ rays in the 120-140 GeV energy range was 
reported~\cite{Bringmann:2012vr}, based on public Fermi 
data~\cite{Atwood:2009ez} (see also~\cite{Boyarsky:2012ca, 
Li:2012qg}). Detailed analysis focusing on the Galactic center 
concluded, with 4.6$\sigma$ C.L.,that there was evidence for 
the $\gamma$-ray line~\cite{Weniger:2012tx}; other works
also confirm the signals~\cite{Tempel:2012ey, Su:2012ft, 
Finkbeiner:2012ez}, though there is room for instrumental 
errors~\cite{Whiteson:2012hr,Hektor:2012ev,Whiteson:2013cs} 
(see also Ref.~\cite{Weniger:2012ms} and references therein).  
Quite recently, the Fermi-LAT Collaboration formally reported 
that the line signals at around $E_{\gamma} = 133$ GeV have 
been detected at 3.3$\sigma$ (1.9$\sigma$) C.L. for local
(global) significance~\cite{Fermi-LAT:2013uma}.

So far, no natural astrophysical models for a source of 
$\gamma$-ray lines with a narrow width, $\Delta E/E < 0.15$, 
have been proposed. Some models (e.g., see 
Ref.~\cite{Aharonian:2012cs} and references therein) try to 
explain the sharp $\gamma$-ray line via the inverse Compton 
scattering of ambient photons by electrons from a neutron star 
in the Klein-Nishina regime; this requires the electrons in the 
wind to be monoenergetic with a small dispersion, less than 
20\%$-$30\%. Hence, the excess of $\gamma$ rays triggered the 
construction of new models of dark matter~\cite{Profumo:2012tr, 
Dudas:2012pb, Cline:2012nw, Choi:2012ap, Kyae:2012vi, 
Lee:2012bq, Acharya:2012dz, Buckley:2012ws, Chu:2012qy, 
Das:2012ys, Kang:2012bq, Feng:2012gs, Yang:2012ha, Park:2012xq,
Huang:2012yf, Hektor:2012jc, Cline:2012bz, Bai:2012qy, Laha:2012fg,
Bergstrom:2012bd, Wang:2012ts, Baek:2012ub, Shakya:2012fj,
Farzan:2012kk, Chalons:2012xf, Asano:2012zv, Rajaraman:2012fu,
Gorbunov:2012sk, Lee:2012ph, Biswas:2013nn, Gu:2013iy, Chen:2013bi,
Endo:2013si, Jackson:2013pjq, Tomar:2013zoa, Toma:2013bka, 
Giacchino:2013bta}\footnote{See also~\cite{Park:2012xq} for 
discussions about morphology differences between the annihilation 
and the decay scenarios.}.

It is challenging to build dark matter models without
conflicting with astrophysical observations. A concern is
the discrepancy between the dark matter annihilation cross
section that yields the correct relic density and the one
that fits the observed $\gamma$-ray signals. Because dark matter
must be electrically neutral, $\gamma$-ray productions from dark
matter annihilations are higher-order processes. For the
annihilation cross section fitting to the 130 GeV $\gamma$ rays,
therefore, tree-level dark matter annihilations lead to an
oversize cross section, in order to account for the measured
relic density.

Another concern is the constraint on continuum $\gamma$-ray
contributions of dark matter annihilation. Dark matters
annihilate into final states (e.g., $W^+ W^-$, $Z^0 Z^0$, 
and so on), and their decay products produce continuum 
photon contributions. The ratio of the number of the continuum
photons, $N_\text{ann}$, to the number of the photons responsible
for the 130 GeV $\gamma$ rays, $N_{\gamma \gamma} + 
N_{\gamma Z}$, is constrained by Fermi data. For example, in 
the case that the dominant annihilation products of 130 GeV 
dark matter are $W^+ W^-$ (or $Z^0 Z^0$), the constraint on 
the ratio is $N_\text{ann}/(N_{\gamma \gamma} + 
N_{\gamma Z}) \lesssim 7$~\cite{Cohen:2012me}.
\footnote{From big bang nucleosynthesis and cosmic microwave 
background anisotropies, we can obtain lower limits on the flux ratio
of the $\gamma$-ray lines to the continuum $\gamma$-rays, 
$\gtrsim 1\times 10^{-3}$, commonly for $W^{+}W^{-}$, 
$b+\bar{b}$, and $e^{+}e^{-}$'s and/or $\gamma$'s 
modes~\cite{Hisano:2011dc}.} 
The eligible dark matter candidate must therefore weakly interact 
with the standard model (SM) particles, except for photons. 
Furthermore, the hadronic contributions of annihilation (decay) 
products from dark matter are also constrained, by the measurement 
of the antiproton-to-proton ratio~\cite{Buchmuller:2012rc, 
Fornengo:2013xda} with PAMELA data~\cite{Adriani:2008zq}.

In this work, we focus on the axion-photon conversion as a source 
of the $\gamma$-ray excess. We propose a scenario as follows: 
energetic axions are produced at the Galactic center, and the axions 
convert into $\gamma$-ray lines through the Primakoff effect in an 
external Galactic magnetic field~\cite{Pirmakoff:1951pj, Sikivie:1983ip}. 
The axion-photon conversion and their oscillating propagation in 
our Galaxy have been extensively studied~\cite{Hooper:2007bq, 
Roncadelli:2008zz, DeAngelis:2008sk, Fairbairn:2009zi, 
Bassan:2010ya, Meyer:2013pny}. We discuss both the conventional 
QCD axions and very light pseudoscalar particles that interact with 
electromagnetic field. The latter are called axionlike particles (ALPs). 
The symbol a in this work refers to both conventional axions and 
ALPs. (See also Refs.~\cite{Conlon:2013txa, Higaki:2013qka, 
Tashiro:2013yea} for cosmological applications.)

This work is organized as follows. First, we briefly review the 
axion-photon conversion and evaluate its probability. Then, we 
evaluate the axion production cross section and the decay rate 
of dark matter to fit the $\gamma$-ray excess. In 
Sec.~\ref{sec:models}, as an illustrative example, we discuss 
supersymmetric axion models in which the 130 GeV $\gamma$-ray 
lines can be fitted. Finally, Sec.~\ref{sec:summary} is devoted to 
summary and discussion.

\section{Axion-photon conversion and its application to the 130 GeV $\gamma$}

\begin{figure}[t!]
\begin{center}
\includegraphics[width=60mm]{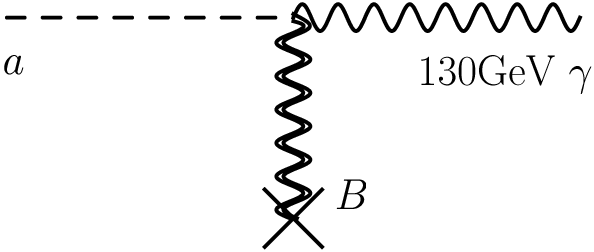}
\caption{Axion-photon conversion in an external magnetic field 
$B$.}
\label{Fig:prima}
\end{center}  
\end{figure}

We consider the scenario that energetic axions are produced at the 
Galactic center and that they convert into $\gamma$ rays in an 
external magnetic field~\cite{Pirmakoff:1951pj} during their
flight to Earth (Fig. 1). In this scenario, the observed $\gamma$-ray
flux $J_\gamma$ is given by
\begin{equation}
\begin{split}
   J_\gamma = P_{a \gamma} \cdot J_a(E_a, \theta, \phi), 
\label{flux}   
\end{split}
\end{equation} 
where $P_{a \gamma}$ is the conversion probability from an 
axion a to a photon, and $J_a(E_a, \theta, \phi)$ is the axion 
flux with energy $E_a$, depending on a direction ($\theta, \phi$). 
In this section, we briefly review the conversion probability and 
write it in a convenient form. Next, we discuss axion production 
and derive the axion production cross section and the decay rate for
fitting the 130 GeV $\gamma$-ray flux.

\subsection{Conversion probability} \label{Sec:conv} 

The conversion process from an axion to a photon is described by 
the following Lagrangian: 
\begin{equation}
\begin{split}
   &\mathcal{L}_{a \to \gamma} = 
   \frac{1}{2}(\partial^\mu a)^2 - \frac{1}{2} m_a^2 a^2 
   - \frac{1}{4} F_{\mu \nu} F^{\mu \nu} 
   - \frac{1}{4} g_{a \gamma} F_{\mu \nu} \tilde F^{\mu \nu} a, 
\end{split}
\end{equation} 
where $F_{\mu \nu}$ is the electromagnetic field strength and 
$\tilde F_{\mu \nu}$ is its dual. Here $g_{a \gamma}$ is the 
effective coupling constant and $m_a$ is the axion mass. 
Although $g_{a \gamma}$ and $m_a$ are related in QCD
axion models, in this work we suppose that they are independent 
of each other. The current upper bound on
$g_{a \gamma}$ is obtained by the CAST experiment, 
$g_{a \gamma} \lesssim 8.8 \times 10^{-11}\text{GeV}^{-1}$ 
for $m_a \lesssim 0.02\text{eV}$~\cite{Arik:2011rx}. 
The conversion probability $P_{a \gamma}$ is given 
by~\cite{Pirmakoff:1951pj, Raffelt:1987im}, 
\begin{equation}
\begin{split}
   P_{a \gamma} 
   &= 
   \bigl| \bigl \langle A(t, x=L) | a(t=0, x=0) \bigr \rangle \bigr|^2
   \\&= 
   \sin^2 2\Theta \sin^2 \Bigl( \frac{1}{2}qL \Bigr),  
\end{split}
\end{equation} 
where $L$ is the distance of the propagation and $\Theta$ is the
mixing parameter of axion and photon,
\begin{equation}
\begin{split}
   \sin\Theta = i \sqrt{\frac{\lambda_+}{\lambda_+ - \lambda_-}}, \ 
   \cos\Theta = i \sqrt{\frac{\lambda_-}{\lambda_+ - \lambda_-}}. 
\end{split}
\end{equation} 
Here $\lambda_\pm$ are the eigenvalues of axion-photon mixing state, 
\begin{equation}
\begin{split}
   \lambda_\pm = \frac{1}{2} \Bigl[ m_a^2 \pm 
   \sqrt{m_a^4 + 4 g_{a \gamma}^2 B^2 E_{\gamma}^2} \ \Bigr] . 
\end{split}
\end{equation} 
With the approximation $E_a \simeq E_\gamma$, the momentum 
transfer $q$ on the conversion is calculated as follows: 
\begin{equation}
\begin{split}
  q = \sqrt{E_{\gamma}^2-\lambda_-} - \sqrt{E_{\gamma}^2-\lambda_+}
  \simeq \frac{1}{2E_{\gamma}} \sqrt{m_a^4 + 4 g_{a \gamma}^2 B^2
    E_{\gamma}^2},
\end{split}
\end{equation} 
in a high-energy limit $E_{\gamma} \gg m_a$. As a result, in the limit of
$E_{\gamma} \gg m_a$, the conversion probability is obtained as follows:
\begin{equation}
\begin{split}
   P_{a \gamma} 
   &= 
   \sin^2 \biggl[ \frac{g_{a \gamma} B L}{2} 
   \sqrt{1+ \Bigl( \frac{m_a^2}{2 g_{a \gamma} B E_\gamma} 
   \Bigr)^2} \biggr] 
   \Bigl[ 1 + \Bigl( \frac{m_a^2}{2 g_{a \gamma} B E_\gamma} 
   \Bigr)^2 \Bigr]^{-1}
   \\&= 
   \sin^2\biggl[ 
   2.4 \times 10
   \Bigl( \frac{L}{7.94\text{kpc}} \Bigr) 
   \sqrt{\Bigl( \frac{130\text{GeV}}{E_{\gamma}} \Bigr)^2 
   \Bigl( \frac{m_a}{10^{-7}\text{eV}} \Bigr)^4 
   +3.2 \Bigl( \frac{g_{a\gamma}}{10^{-10}\text{GeV}^{-1}} \Bigr)^2
   \Bigl( \frac{B}{10\mu\text{G}} \Bigr)^2}  \ 
   \biggr] 
   \\& ~~~ \times
   \Bigl[ 
   1+0.31 \Bigl( \frac{m_a}{10^{-7}\text{eV}} \Bigr)^4 
   \Bigl( \frac{10^{-10}\text{GeV}^{-1}}{g_{a\gamma}} \Bigr)^2 
   \Bigl( \frac{10 \mu \text{G}}{B} \Bigr)^2 
   \Bigl( \frac{130 \text{GeV}}{E_{\gamma}} \Bigr)^2 
   \Bigr]^{-1}. 
\label{Eq:P_agamma}   
\end{split}
\end{equation}

\begin{figure}[h!]
\begin{center}
\includegraphics[width=90mm]{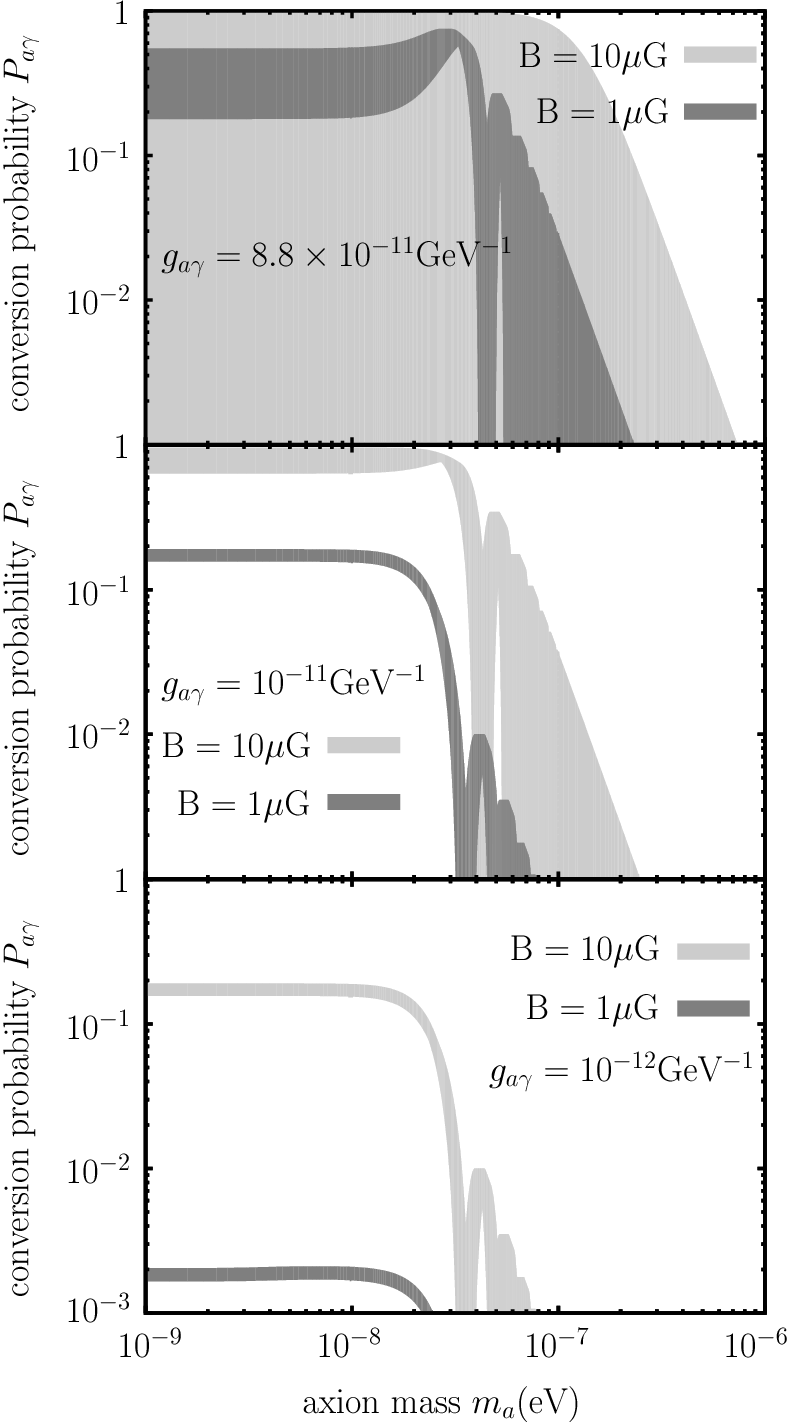}
\caption{Conversion probability in the magnetic field, 
  $B = 1 \mu$G, and $10 \mu$G.  The width of each band 
  represents ambiguities of both the energy of observed 
  $\gamma$-rays and the distance from the Galactic center 
  to the solar system. From the top to the bottom, we took 
  the coupling constant to be $g_{a \gamma} = 8.8 \times 
  10^{-11}$GeV$^{-1}$, $1 \times 10^{-11}$GeV$^{-1}$, 
  and $1 \times 10^{-12}$GeV$^{-1}$, respectively. }
\label{Fig:conv}
\end{center}  
\end{figure}

In Fig.~\ref{Fig:conv} we plot $P_{a \gamma}$ as a function 
of $m_a$. We take $g_{a \gamma} = 8.8 \times 
10^{-11}$GeV$^{-1}$ in the top panel, which is the upper
bound of $g_{a \gamma}$, $g_{a \gamma}=1 \times 
10^{-11}$GeV$^{-1}$ in the middle panel, and 
$g_{a \gamma} = 1 \times 10^{-12}$GeV$^{-1}$ in the 
bottom panel. In each panel, the magnetic field in the Galaxy 
is assumed to be a uniform distribution, $B = 10\mu$G 
(bright band) and $B = 1\mu$G (dark band). The width of 
each band represents the ambiguities of both the energies 
of the $\gamma$-ray $E_\gamma = 129.8 \pm 
2.4^{+7}_{-13}$GeV~\cite{Weniger:2012tx}, and the 
distance from the Galactic center to the solar system, 
$L = 7.94 \pm 0.42$kpc~\cite{Eisenhauer:2003he}.

In the top panel, the probability for $B=10\mu$G spans a
broad range. This is because the argument of sine in 
Eq.~(\ref{Eq:P_agamma}) is $\mathcal{O}(10)$ for 
$g_{a \gamma} = 8.8 \times 10^{-11}$GeV$^{-1}$ and 
$B=10\mu$G. Thus, $P_{a \gamma}$ is highly sensitive to 
tiny variations in $E_\gamma$ and/or $L$ within their ambiguities.

With the exception of the case where $B=10\mu$G and 
$g_{a \gamma} = 8.8 \times 10^{-11}$GeV$^{-1}$, the 
probability is constant with $m_a$ for the fixed $B$ and 
$g_{a \gamma}$ for lower mass regions. This is understood 
from Eq.~(\ref{Eq:P_agamma}). 
In Eq.~(\ref{Eq:P_agamma}), the term proportional to 
$B^2$ in the square root in the second line is larger than the
term proportional to $m_a^4$, and the third line is unity. 
On the other hand, for larger mass regions, the terms 
containing $m_a$ are dominant on both the second and 
third lines in Eq.~(\ref{Eq:P_agamma}). The third line 
suppresses $P_{a \gamma}$ by a factor of $m_a^{-4}$, 
which gives oscillation damping as shown in Fig.~\ref{Fig:conv}.

Before closing this subsection we would like to mention magnetic 
field distributions in the Galaxy and magnetic field strength near 
the Galactic center. Because realistic magnetic field distributions 
remain a matter of research, we adopted a uniform distribution 
as a practical approximation. Another model for the distributions 
is the turbulent Galactic magnetic field, which is of constant 
magnitude and has a random direction in each small patchy 
domain~\cite{Mirizzi:2007hr}. In this model, the conversion 
probability per single domain is
\begin{equation}
\begin{split}
   P_0 \simeq \frac{g_{a \gamma}^2 \langle |B|^2 \rangle s^2}{4} 
   \frac{\sin^2 \bigl( \pi s/l_0 \bigr)}{\bigl( \pi s/l_0 \bigr)^2}, 
\end{split}
\end{equation} 
where $s$ is typical size of domains and $l_0 = 4 \pi E_\gamma/m_a^2$ 
is the oscillation length. In the limit of $N P_0 \ll 1$ ($N$ is the number of 
domains), assuming the travelling distance $L$ is much larger than $s$, 
the conversion probability is 
\begin{equation}
\begin{split}
   P_{a \gamma} 
   &= 
   \frac{1}{3} \Bigl[ 1- \exp\Bigl(-\frac{3P_0 L}{2s} \Bigr) \Bigr] 
   \\&= 
   \frac{1}{3} \biggl\{ 1 - \exp
   \biggl[ -3.85 \times 10^4 
   \Bigl( \frac{g_{a \gamma}}{10^{-11}\text{GeV}^{-1}} \Bigr)^2 
   \Bigl( \frac{B}{10\mu\text{G}} \Bigr)^2 
   \Bigl( \frac{E}{130\text{GeV}} \Bigr)^2 
   \Bigl( \frac{10^{-7} \text{eV}}{m_a} \Bigr)^4 
   \\& ~~ \times 
   \Bigl( \frac{0.01\text{pc}}{s} \Bigr)
   \Bigl( \frac{L}{7.94\text{kpc}} \Bigr) 
   \sin^2 \Bigl[ 3.0 \times 10^{-3} 
   \Bigl( \frac{s}{0.01\text{pc}} \Bigr)
   \Bigl( \frac{130\text{GeV}}{E_\gamma} \Bigr) 
   \Bigl( \frac{m_a}{10^{-7}\text{eV}} \Bigr)^2 
   \Bigr]
   \biggr]
   \biggr\}. 
\end{split}
\end{equation} 
Assuming $s = 0.01\text{pc}$, for larger $g_{a \gamma}$, 
the conversion probability in the turbulent magnetic field is 
almost constant, and is close in value to the one in the uniform 
distribution. For a smaller coupling, $g_{a \gamma} \lesssim 
10^{-11}$, assuming the same size domain, the conversion 
probability in the turbulent distribution is smaller than that in 
the uniform distribution by one or two orders.

Although the distributions are left to be considered, the 
large-scale magnetic field strength is steadily developing, and 
is indicated as $\sim 10 \mu\text{G}$. On the other hand, 
near the Galactic center, the strength can reach very high values,
$\mathcal{O}(100\mu\text{G})$~\cite{Beck:2013bxa}. 
Under such strong magnetic fields, as is given in 
Eq.~\eqref{Eq:P_agamma}, the conversion probability highly 
oscillates as a function of traveling distance $L$ and the effective
coupling $g_{a \gamma}$, and can be $\sim 1$. However, 
in such a situation, high-energy photons also have high 
probabilities of converting to axions. The flux ratio of axions and 
$\gamma$ rays is therefore expected to be not very different 
with or without a very strong magnetic field near the Galactic center. 
Furthermore, there is no certain interpretation of the strength near 
the Galactic center, which possesses large uncertainties that depend 
on the observational data used in the analyses. Taking these facts 
into account, for our present purpose of showing that $\sim 100 
\text{GeV } \gamma$ rays from the Galactic center are produced 
by the axion-photon conversion, it is acceptable way to use a uniform 
distribution with strength $\sim 10\mu\text{G}$.

\subsection{Possible scenarios to fit the observed $\gamma$-ray line}  
\label{sec:line} %

\begin{figure}[t!]
\begin{center}
\includegraphics[width=100mm]{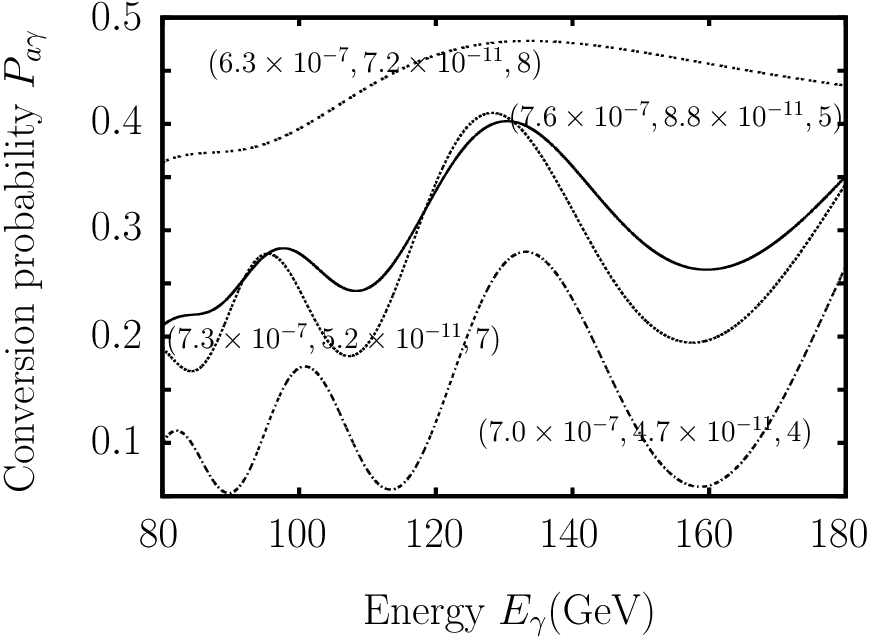}
\caption{Conversion probabilities as a function of $E_{\gamma}$.
  Values on each line denote ($m_a$(eV), 
  $g_{a \gamma}$(GeV$^{-1}$), $B$($\mu$G)). }
\label{Fig:osci}
\end{center}  
\end{figure}

Here we show scenarios explaining the 130 GeV $\gamma$-ray
lines through the use of the axion-photon conversion. The 
conversion probability is an oscillating function of $E_\gamma$ 
[see Eq.~(\ref{Eq:P_agamma})]. In Fig.~\ref{Fig:osci} we 
plot the conversion probability as a function of $E_\gamma$ 
for some parameter sets of ($m_a$, $g_{a \gamma}$, $B$). 
We find that observed bump shapes of $\gamma$ rays are 
produced by the oscillation behavior for some appropriate 
parameter sets even if the energy of the axions is not tuned 
to 130 GeV. Therefore we have at least two types of scenarios 
to fit the observational data: 
\begin{itemize}
\item Monochromatic axions with the energy of $\sim$ 130 
GeV need to be produced. Expected parameters are a large
effective coupling, $g_{a \gamma} \gtrsim 10^{-12} 
\text{GeV}^{-1}$, and small axion mass $m_a \lesssim 
10^{-7}\text{eV}$  (see Fig.~\ref{Fig:conv}).
\item It is not necessary for the axions to be monochromatic. 
There needs to be a fine-tuning in the sets of parameters 
($m_a$, $g_{a \gamma}$, $B$) with a cutoff energy
($\lesssim$ 200 GeV) so that a bump shape is produced
(see Fig.~\ref{Fig:osci}).
\end{itemize}

In this work we focus on the first scenario. An attractive
candidate of the source for the monochromatic axion is a
decaying or an annihilating dark matter. Thus the search for
origins of the 130 GeV $\gamma$ rays can be a bridge to the new
physics that predicts dark matter. The second scenario will
be discussed in a separate paper.

\subsection{Axion production} 

The simplest production of energetic axions is by the decay 
processes of long-lived heavy particles, $\psi$. At the lowest 
order, the process is $\psi \to a \psi'$. Here $\psi'$ is a stable
particle, and we have assumed a $Z_2$-parity conservation. 
Under this parity, SM particles have a plus charge and $\psi$ 
and $\psi'$ have a minus charge. The $Z_2$-parity 
conservation makes $\psi'$ ($\psi$) stable (long-lived), and, 
hence, it is a dark matter candidate [cf., a lightest neutralino 
in supersymmetry (SUSY) models with R-parity conservation, 
a Kaluza-Klein photon in universal extra dimension models 
with KK-parity conservation, and so on]. 
We introduce a symbol $\langle \Gamma \rangle_{a}$ as 
the partial decay width of the above process. In addition to the 
tree-level process, we need to draw attention to higher-order 
processes, which are associated with fermions in final states, 
$\psi \to a \psi' f \bar f$. Observations of antiprotons and 
continuum $\gamma$ rays limit the production rates of these 
fermions. We discuss this issue later.

Another production of energetic axions is by annihilation
processes of heavy neutral particles. These processes are
\begin{itemize}
\item $\psi \psi \to a a$
\item $\psi \psi \to a Z$
\item $\psi \psi \to a h$
\end{itemize} 
We introduce the symbols $\langle \sigma v \rangle_{aa}$, 
$\langle \sigma v \rangle_{aZ}$, and $\langle \sigma v 
\rangle_{ah}$ as the cross sections of these processes. The $Z$ 
boson and the Higgs boson in these processes decay into charged 
particles, but the production rates are limited by cosmic ray
observations. We discuss this issue later.

In the next subsection, we evaluate decay rates and annihilation 
cross sections for the axion production to fit the observed 
$\gamma$ rays.

\subsection{Partial cross sections or decay rates fitted by observations}

The most important element of this work is the proposal to fit the 
$\mathcal{O}(100\text{GeV})$ $\gamma$-ray excess from the 
Galactic center without conflicting with cosmic ray observations. 
To do this, we show that our scenario can provide the observed
$\gamma$-ray flux through the axion-photon conversion with the 
appropriate axion production cross sections and decay rate. 
The axion production cross section and decay rate to fit the 
$\gamma$-ray excess via the axion-photon conversion are expressed by
\begin{equation}
\begin{split}
   \langle \sigma v \rangle_{aa(aZ,ah)} &= 
   c_\gamma \langle \sigma v \rangle_{\gamma \gamma} 
   / P_{a \gamma}^\text{mean}, 
   \\[1mm]
   \langle \Gamma \rangle_a &= 
   \langle \Gamma \rangle_\gamma 
   / P_{a \gamma}^\text{mean}, 
\label{Eq:dec/ann}   
\end{split}
\end{equation} 
where $\langle \sigma v \rangle_{\gamma \gamma}$ and 
$\langle \Gamma \rangle_{\gamma}$ denote the 
$\gamma$-ray production cross section and decay rate that fit 
the observed 130 GeV $\gamma$ rays.
Here $c_{\gamma} = 1$ for $\langle \sigma v \rangle_{aa}$ and 
$c_{\gamma} = 1/2$ for $\langle \sigma v \rangle_{aZ(ah)}$. 
The symbol $P_{a \gamma}^\text{mean}$ is a mean value of the 
conversion probability, which is defined as follows: 
\begin{equation}
\begin{split}
   P_{a \gamma}^\text{mean}
   \equiv 
   \frac{P_{a \gamma}^\text{max} + 
   P_{a \gamma}^\text{min}}{2}.  
\label{Eq:mean1}   
\end{split}
\end{equation} 
Here $P_{a \gamma}^\text{max} (P_{a \gamma}^\text{min})$ 
is the maximal (minimal) conversion probability within the ranges 
of $E_\gamma = 129.8 \pm 2.4_{-13}^{+7}\text{GeV}$ and 
$L = 7.94 \pm 0.42$kpc~\cite{Eisenhauer:2003he}. 
In this calculation, we do not include the effects of conversions from 
$\gamma$ rays to axions or reconversion processes $a \to \gamma 
\to a$. We comment about these issues in the summary and discussion.

In order to parameterize $P_{a \gamma}$, we introduce a parameter
$\tilde m_a$, which is a border point that separates the region with
regard to $m_a$ dependence. For $m_a \lesssim \tilde m_a$, 
$P_{a \gamma}$ is almost independent of $m_a$, and the mean 
value is straightforwardly obtained from Eq.~\eqref{Eq:mean1}. 
We introduce a symbol $\tilde P_{a \gamma}^\text{mean}$, 
which is $P_{a \gamma}^\text{mean}$ in the region $m_a \lesssim 
\tilde m_a$. For the oscillation damping regime, 
$P_{a \gamma}^\text{mean}$ is parameterized as follows:
\begin{equation}
\begin{split}
   P_{a \gamma}^\text{mean}
   = \tilde P_{a \gamma}^\text{mean}
   \Bigl( \frac{m_a}{\tilde m_a} \Bigr)^{-4} 
   ~~~ (\text{For } m_a \gtrsim \tilde m_a ). 
\label{Eq:mean2}   
\end{split}
\end{equation} 
Here $\tilde P_{a \gamma}^\text{mean}$ and $\tilde m_a$ for each 
parameter set are listed in the fourth column of Table~\ref{Tab:Pmean} 
as coefficients of $(m_a/\tilde m_a)^{-4}$, with the denominator in 
parentheses.

\begin{table}
\begin{center}
\caption{Mean values of conversion probabilities 
$P_{a \gamma}^\text{mean}$. Symbol $\tilde m_a$ separates 
the region in regard to $m_a$ dependence, and $\tilde m_a$ for 
each parameter sets are given in the denominator in parentheses 
in fourth column. 
\label{table:input}}
\vspace{1mm}
\begin{tabular}{llll} \hline
~ $g_{a \gamma}$
& $B$ 
& $P_{a \gamma}^\text{mean}$ for $m_a \lesssim \tilde m_a$ ~ 
& $P_{a \gamma}^\text{mean}$ for $m_a \gtrsim \tilde m_a$ ~   
\\[0.7mm] \hline
~$8.8 \times 10^{-11}$ ~
& $1\mu$G
& $3.71 \times 10^{-1}$
& $3.71 \times 10^{-1} (m_a/3.70 \times 10^{-8}\text{eV})^{-4}$ ~ 
\\[0.7mm] 

&$10\mu$G ~
& $5.00 \times 10^{-1}$
& $5.00 \times 10^{-1} (m_a/1.27 \times 10^{-7}\text{eV})^{-4}$ ~ 
\\[0.7mm] \hline
~$1 \times 10^{-11}$ ~  
& $1\mu$G        
& $1.72 \times 10^{-1}$
& $1.72 \times 10^{-1} (m_a/1.73 \times 10^{-8}\text{eV})^{-4}$
\\[0.7mm] 
           
& $10\mu$G
& $8.02 \times 10^{-1}$  
& $8.02 \times 10^{-1} (m_a/3.63 \times 10^{-8}\text{eV})^{-4}$
\\[0.7mm] \hline
~$1 \times 10^{-12}$
&$1\mu$G
& $1.88 \times 10^{-3}$
& $1.88 \times 10^{-3} (m_a/1.70 \times 10^{-8}\text{eV})^{-4}$
\\[0.7mm] 

&$10\mu$G
& $1.72 \times 10^{-1}$
& $1.72 \times 10^{-1} (m_a/3.63 \times 10^{-8}\text{eV})^{-4}$
\\[0.7mm] \hline
\label{Tab:Pmean}
\end{tabular}
\end{center}
\end{table}

The annihilation cross section into two $\gamma$ lines to fit the
130 GeV $\gamma$-ray excess is 
$\langle \sigma v \rangle_{\gamma \gamma} = (1.27 \pm 
0.32^{+0.18}_{-0.28}) \times 10^{-27} \text{cm}^3 {\rm s}^{-1}$ 
with its mass $m_{\text{DM}} = 129.8 \pm 2.4^{+7}_{-13}$GeV 
in the case of the Einasto profile~\cite{Weniger:2012tx}. 
Similarly, the best-fit
lifetime for decaying into $\gamma$ rays is $\tau_\gamma = 
1.24^{+1.20}_{-0.44} \times 10^{28}$~s~\cite{Buchmuller:2012rc}, 
with a similar setup as in Ref.~\cite{Weniger:2012tx}.

\begin{figure}[t!]
\begin{center}
\includegraphics[width=100mm]{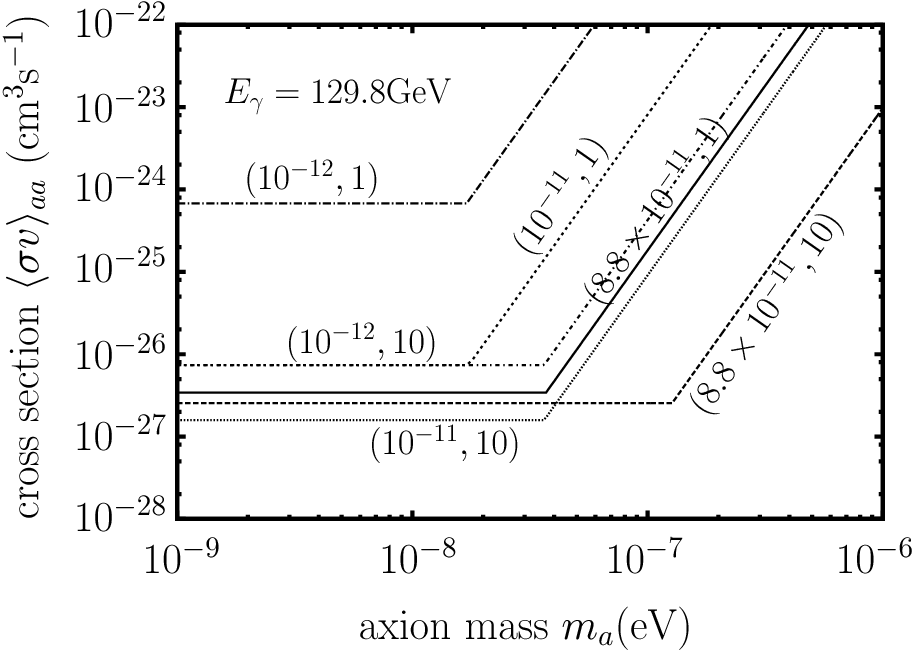}
\caption{Annihilation cross section into axions to fit the 130GeV 
  $\gamma$-ray measurements. Each label attached on the line 
  shows the parameter set of $g_{a \gamma}$ in unit of GeV$^{-1}$ 
  and B in unit of $\mu \text{G}$, which are the same as those 
  used in Fig.~\ref{Fig:conv}. }
\label{Fig:sigmav}
\end{center}  
\begin{center}
\includegraphics[width=100mm]{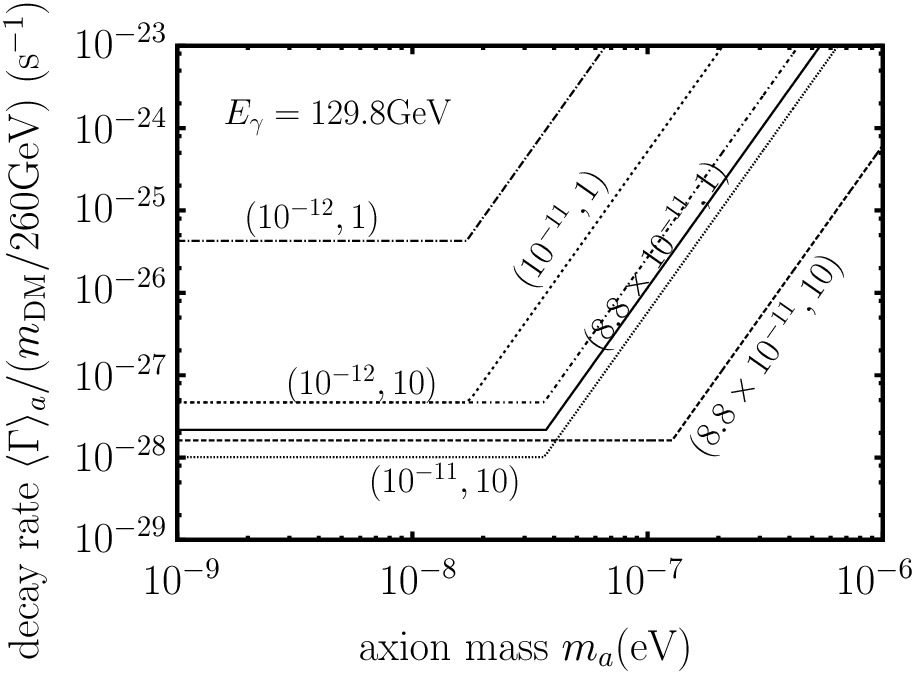}
\caption{Same as Fig.\ref{Fig:sigmav}, but for the partial decay rate
  into axion.}
\label{Fig:Gamma}
\end{center}  
\end{figure}

The axion production cross sections $\langle \sigma v \rangle_{aa 
(aZ,ah)}$ and partial decay rate $\langle \Gamma \rangle_a$ to fit 
the $\gamma$-ray excess are evaluated from 
$P_{a \gamma}^\text{mean}$ listed in Table~\ref{Tab:Pmean} 
and Eq.~(\ref{Eq:dec/ann}). The results are plotted in 
Figs.~\ref{Fig:sigmav} and \ref{Fig:Gamma}, respectively. 
In both plots we take the same reference values for $g_{a \gamma}$ 
and $B$ as the ones used in Fig.~\ref{Fig:conv}.\footnote{When 
we suppose turbulent magnetic field distributions,
larger annihilation cross sections and decay rates are required
to fit the $\gamma$-ray spectrum for smaller couplings, 
$g_{a \gamma} \lesssim 10^{-11}$.
Production cross sections and decay rates that are too large,
however, may conflict with cosmic ray constraints, e.g., continuum
$\gamma$-ray contributions. Thus, if turbulent magnetic field
distributions are realized, larger $g_{a \gamma}$ would be expected, 
in order to explain the $\gamma$-ray excess.}
Smaller $g_{a \gamma}$ and weaker $B$ lead to smaller $P_{a 
\gamma}$; hence, $\langle \sigma v \rangle_{aa (aZ,ah)}$ 
and $\langle \Gamma \rangle_a$ need to be larger. An
exception is the case of the parameter $(8.8 \times 10^{-11}, 
10)$. As is explained in Sec.~\ref{Sec:conv}, in this case, 
$P_{a \gamma}$ spans a broad range, and hence its mean 
value is smaller than that of the parameter $(10^{-11}, 10)$.

In some regions in Fig.~\ref{Fig:sigmav}, $\langle \sigma v
\rangle_{aa}$ is larger than the canonical annihilation cross 
section, which fits the correct relic abundance in thermal relic 
scenarios, $\langle \sigma v \rangle \sim 2.2 \times 10^{-26} 
\text{cm}^3 {\rm s}^{-1}$ (e.g., see~\cite{Steigman:2012nb}).  
Hence, in these regions, we need nonthermal production of dark 
matter, e.g., through decays of another massive particle, or a large 
boost factor for the dark matter annihilation cross section at the Galactic
center in the current Universe.

Before closing this section, we mention the constraints
on the axion production cross section. 
In axion production, annihilation modes $\psi \psi \to aa Z(h)$ 
can cause anomalous excesses of antiproton flux and continuum 
$\gamma$-ray flux from the secondary products of $Z$ and $h$. 
From the PAMELA measurements for the antiproton-to-proton 
flux ratio, a constraint on the annihilation modes into $Z$ bosons 
is obtained, $\langle \sigma v \rangle_{Z} \lesssim 10^{-25} 
\text{cm}^3 {\rm s}^{-1}$ in the MED propagation
model~\cite{Fornengo:2013xda}. 
The comparable constraint is obtained from the continuum 
$\gamma$-ray measurements, $\langle \sigma v \rangle_{Z} 
\lesssim 10^{-25} \text{cm}^3 
{\rm s}^{-1}$~\cite{Buchmuller:2012rc, Cohen:2012me}. 
Since the annihilation modes are three-body final states, in general, 
the cross section is smaller than that for $\psi \psi \to aa$ by at 
least $\mathcal{O} (\alpha)$. Here, $\alpha$ is the fine-structure 
constant. Thus, from the relation $\langle \sigma v 
\rangle_{aaZ(h)}/\langle \sigma v \rangle_{aa} \lesssim 10^{-2}$,
we can evaluate the bound on the axion production cross section, 
$\langle \sigma v \rangle_{aa} \lesssim 10^{-23} \text{cm}^3 
{\rm s}^{-1}$. 
Note that constraints on the axion single production, $\psi \psi \to 
a Z(h)$, are more severe than those for the axion pair productions. 
The bound on the cross section is $\langle \sigma v \rangle_{aZ} 
\lesssim 10^{-25} \text{cm}^3 {\rm s}^{-1}$.

In axion production from the decay process, similarly, the partial 
lifetimes of the modes $\psi \to a \psi' Z$, $\psi \to a \psi' h$, 
and $\psi \to a \psi' f \bar f$ are constrained by the measurements 
of the antiproton and the continuum $\gamma$ ray, 
$1/\langle \Gamma \rangle_{aZ(h)} \gtrsim 
10^{27\text{-}28}$s~\cite{Fornengo:2013xda, Garny:2012vt}. 
With the same discussion as for the annihilation cross sections, the 
bound on the partial lifetime is $1/\langle \Gamma \rangle_a \gtrsim 
10^{25\text{-}26}$s.

\section{Models}  \label{sec:models} 

In this section, as an illustrative example, we discuss two dark matter 
models with regard to axion productions. Our purpose here is just to 
show that some dark matter models surely produce monochromatic 
axions without producing problematic cosmic rays.

\subsection{Axion production from decay of the long-lived dark matter} 

\begin{figure}[t!]
\begin{center}
\includegraphics[width=60mm]{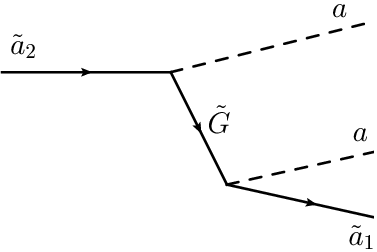}
\caption{Decay process producing axion.}
\label{Fig:decay}
\end{center}  
\end{figure}

As we have seen in Fig.~\ref{Fig:sigmav}, a preferred lifetime 
of the parent dark matter is around $10^{25} - 10^{28}$ s. 
In Ref.~\cite{Arvanitaki:2008hq} it is pointed out that such a 
long lifetime can be achieved for the process suppressed by 
$1/M_{\rm GUT}^4$, namely, the process mediated by the 
GUT-scale-suppressed dimension-six operators. If such a 
suppressed interaction can produce axions, our scenario can be 
realized. On the other hand, a dimension-five GUT-scale-suppressed 
interaction leads to the much shorter lifetime, $\mathcal{O}$(1 s).

One of the natural classes of models that predict dark matter is 
supersymmetry, and it is worth considering the possibility of realizing 
our scenario in the SUSY model. 
The lightest SUSY particle (LSP) with conserved R parity is stable. 
SUSY models sometimes predict a long-lived particle that can decay 
into LSP. Moreover, the SUSY should be extended to gravity sector. 
The superpartner of graviton, gravitino $\tilde{\psi}$, interacts 
with all particles, with the interaction suppressed by $1/M_{pl}$. 
Here $M_{pl} = 2.44 \times 10^{18}$ GeV is the reduced Planck mass. 
Another suppressed coupling of this theory is 
axino($\tilde{a}$)-gaugino-$\gamma$($Z$) coupling, the 
suppression factor of which is $g_{a \gamma}$ . It should be of 
the order of $10^{-10}$$-$$10^{-12}$GeV$^{-1}$ in our scenario.

Both of the interactions above are five-dimensional operators and do 
not directly induce very long-lived particles of our interest. Therefore, 
we introduce two axion and axino pairs, ($a_1$, $\tilde{a}_1$) and 
($a_2$, $\tilde{a}_2$). 
The decay $\tilde{a}_2 \rightarrow a_1  a_2 \tilde{a}_1$ can be 
mediated by a gravitino (Fig.~\ref{Fig:decay}). It is suppressed 
by $1/M_{pl}^4$ and may be long enough to be of interest to us. 
The axion energy can be monochromatic if the mass difference 
between $a_2$ and the gravitino is small, so that $a_2$ has low 
energy and, hence, the kinematics of $\psi^*_G$(virtual gravitino) 
$\rightarrow a_1 \tilde{a}_1$ is approximately two body.
In that case, however, we also have to worry about $\tilde{a}_2 
\to 2\gamma \tilde{a}_1$ being mediated by a gaugino exchange, 
which is proportional to $(g_{a_1 \gamma} g_{a_2 \gamma})^2$.  
The suppression of the gaugino exchange decay can be achieved if 
$g_{a_2 \gamma}$ is suppressed. This can be achieved if $a_2$ 
belongs to a Peccei-Quinn sector that does not directly couple to a 
SM gauge boson. The decay rate of $\tilde{a}_2 \to 
\tilde{a}_1 a_1 a_2$ can be expressed as
\begin{equation}
\begin{split}
   \Gamma(\tilde a_2 \to \tilde a_1 + a_1 + a_2) 
   &  \simeq 
   \left( \hspace{-0.8mm} \frac{1}{\sqrt{2} M_{pl}} \hspace{-0.8mm} \right)^4 
   \left( \frac{m_{\tilde a_2} + m_{\tilde G}}
   {m_{\tilde a_2}^2 - m_{\tilde G}^2} \right)^2 
   \left( m_{\tilde a_2} - m_{\tilde a_1} \right)^7
   \\&  \simeq 
   3.4 \times 10^{-27}[\text{s}^{-1}] 
   \left( \hspace{-0.6mm} \frac{1\text{MeV}}
   {m_{\tilde G} - m_{\tilde a_2}} \hspace{-0.6mm} \right)^2 
   \left( \hspace{-0.6mm} \frac{m_{\tilde a_2} - m_{\tilde a_1}}
   {260\text{GeV}} \hspace{-0.6mm}
   \right)^7. 
\end{split}
\end{equation} 
Thus, in the reference scenario, although tight degeneracy in mass 
between $\tilde G$ and $\tilde a_2$ is required, monochromatic
axions can be successfully produced within the favored range of the 
lifetime.

\subsection{Axion production from annihilation} 

For the second example, we assume that a neutralino $\tilde \chi^0$ 
is the LSP and is the dark matter. Under R-parity conservation, axions 
are produced by the neutralino annihilation,
$\tilde \chi^0 \tilde \chi^0 \to aa$, via saxion s-channel exchange.

The annihilation process $\tilde \chi^0 \tilde \chi^0 \to aa$ provides 
the largest cross section when $m_s^2 \simeq q^2$, i.e., $m_s \simeq 
2m_{\tilde \chi^0}$, and hence the width part dominates over the 
part of momentum transfer in the saxion propagator,
\begin{equation}
\begin{split}
   \sigma v (\tilde \chi^0 \tilde \chi^0 \to aa) 
   &= 
   \left( \frac{\alpha_k C_k}{4 \pi F_a} \right)^2 
   \left( \hspace{-0.4mm} \frac{1}{\sqrt{2} F_a} \hspace{-0.4mm} \right)^2 
   \frac{1}{(q^2 - m_s^2)^2 + (m_s \Gamma_s)^2} m_{\tilde \chi^0}^6
   \\[1mm]& \simeq 
   3.5 \times 10^{-28} [\text{cm}^3 \text{s}^{-1}] 
   \Bigl( \frac{130\text{GeV}}{m_{\tilde \chi^0}} \Bigr)^2 
   \Bigl( \frac{\alpha_k C_k}{10^{-2}} \Bigr)^2 ,
\end{split}
\end{equation} 
where $F_a$ is the axion decay constant and $\alpha_k = g_k^2/4\pi$ 
[$g_k$ is the coupling constant for $U(1)_Y$ or the $SU(2)_L$ gauge 
group]. The constant $C_k$ is the model-dependent parameter with 
${\cal O}$(1) or smaller. The interaction of neutralinos and the saxion 
may be modified after SUSY breaking. The constant $C_k$ also includes 
this uncertainty. Here we assume that the dominant decay channel of 
the saxion is $s \to aa$; then, the decay width is $\Gamma_s \simeq 
m_s^3/2F_a^2 \simeq 4m_{\tilde \chi}^3/F_a^2$.

As shown in Fig.~\ref{Fig:sigmav}, the cross section for the axion 
production is required to be larger than $10^{-27} \text{cm}^3
\text{s}^{-1}$. For reproducing the 130 GeV $\gamma$ rays, 
therefore, this type of SUSY model needs a boost factor of 
$\mathcal{O}(1)$ in addition to the relation $m_s \simeq 
2m_{\tilde \chi^0}$.

We currently need more precise studies of decay or annihilation 
processes, which should include information of dark matter profiles 
of the Galaxy. Other possible decay or annihilation processes of axion 
productions should be also studied extensively. We will discuss these 
issues in detail in a separate publication~\cite{yama}.

\section{Summary and discussion}  \label{sec:summary}

An excess of $\gamma$-ray lines in the energy range 120-140 GeV 
was found in public Fermi data~\cite{Bringmann:2012vr}. We have
proposed a scenario such that the excess is explained through the 
axion-photon conversion. We have shown that the $\gamma$ rays 
produced via the axion-photon conversion can fit the observed 
narrow-line spectrum feature with the axion mass $m_a \lesssim 
10^{-6}$~eV, the Galactic magnetic field $B \sim 1$-$10 \mu 
\text{G}$, and the axion-photon coupling constant 
$g_{a\gamma} \lesssim 8.8 \times 10^{-11} {\rm GeV}^{-1}$.

We have evaluated the axion production cross section and the 
decay rate of heavy neutral particles for successfully fitting the 
130 GeV $\gamma$ rays. We have shown example models in 
particle physics, one of which is the decaying axino scenario, 
and the other is the annihilating neutralino scenario. In both 
scenarios, a narrow-line spectrum of axions can be produced 
with the required cross section and decay rate.

In this work we have focused on the monoenergetic axion 
productions from annihilations and decays of heavy particles. 
The oscillation behavior of the conversion probability, however, 
must be considered as important. In fact, a broader spectrum of 
axions can fit the observed 130 GeV $\gamma$ rays, because 
the oscillation dependence of the conversion probability will make 
a bump structure with tuning parameters (see Fig.~\ref{Fig:osci}).

The scenario we propose here has attractive points. First, in our scenario, 
the $\gamma$-ray line is indirectly produced by the axion-photon 
conversion. In the axion production, without complicated gimmicks, 
productions of problematic cosmic rays (e.g., continuum $\gamma$ rays, 
antiprotons, and so on) are sufficiently suppressed. Second, the scenario 
is testable in future experiments and observations. Expected axion-photon 
coupling with the goal of fitting the $\gamma$-ray excess is just below the 
current bound from the CAST experiment~\cite{Arik:2011rx}. 
Next-generation axion search experiments are planned to reach the favored 
values~\cite{Irastorza:2011gs, Graham:2011qk}.  
Future $\gamma$-ray observations, H.E.S.S. II~\cite{phaseII}, 
Cherenkov Telescope Array (CTA)~\cite{Consortium:2010bc}, and 
GAMMA-400~\cite{Galper:2012ji}, will also observe the 130 GeV 
lines with much better sensitivities. We can discriminate our scenario 
from other models by using line shapes and the emission profile 
(morphology) of the spatial distribution~\cite{Bergstrom:2012vd, 
Weniger:2013tza}. Third, our scenario would provide a probe not 
only into the nature of dark matter, but also into models behind the 
axion. In such models, axion properties are occasionally described by 
high-scale parameters, which are beyond the energy scale of collider 
experiments~\cite{Linde:1987bx, Hertzberg:2008wr}. Combining 
observations of $\gamma$ rays with experimental results of the 
axion search will thereby shed light on the energy scale, interactions 
with axions and the dark matter, and so on.

Several points have to be carefully researched to test the scenario in 
future experiments/observations, and for the scenario to be a more 
sensitive probe to the nature of dark matter. First, dark matter 
distributions need detailed treatments. Indeed, the evaluation of 
$\gamma$-ray flux needs the spatial integral over the line of sight with 
dark matter profiles. However, there are still large uncertainties in both 
the distance from the Galactic center to the Earth and spatial distributions 
of the dark matter profiles. One important thing we know for certain is 
that dark matter density steeply damps outside of Galactic center region. 
Thus, as a practicable approximation, this work supposes a uniformly
distributing profile localizing near the Galactic center. With improved 
understanding of the profiles, reevaluation of the flux should be carried out.
Second, we comment on the reconversion processes $a \to \gamma \to a$. 
Large conversion probabilities will modify the observed $\gamma$-ray 
flux $J_\gamma$ via the axion-photon conversion, $J_\gamma \simeq 
(P_{a \gamma} - P_{a \gamma}^2 + \cdot \hspace{-1.5pt} \cdot 
\hspace{-1.5pt} \cdot) \cdot J_a$, where $J_a$ is axion flux, and the 
axion production cross section or decay rate is expected to be larger than 
that without the reconversion processes by, at most, a factor of a few. 
So far, however, such modifications could not have been distinguished 
by any observations of $\mathcal{O}(100\text{GeV}) \ \gamma$-ray 
\cite{Atwood:2009ez, Fermi-LAT:2013uma}.
With the current accuracy of observations, the modification does not 
change our conclusion that high-energy $\gamma$-line excess from 
the Galactic center can be interpreted by the axion-photon conversion 
without conflicting cosmic ray observations. However, in future, 
improvement of the observations will require precise calculation, taking 
into account the reconversion to confirm the scenario and determine 
axion parameters. 
Third, we mention the inverse conversion processes, $\gamma \to a$. 
Since the conversion probability in some benchmark points is large 
(Table~\ref{Tab:Pmean}), the $\gamma$-ray flux from distant sources 
may be attenuated by the reconversion. The attenuation can provide the 
observed large degree of transparency of the universe to $\gamma$ 
rays~\cite{Aharonian:2005gh}. Future $\gamma$-ray observations 
will confirm our scenario by observing the predicted $\gamma$-ray flux 
from various sources, and by checking the dependence of the energy 
and traveling distance on the conversion probability. We will discuss
these issues in detail in a separate publication~\cite{yama}.


\section*{Acknowledgments}

K.K. thanks K. Nakayama and O. Seto for useful discussions. This work
is supported by the Grant-in-Aid for Scientific research
from the Ministry of Education, Science, Sports, and Culture, Japan,
Nos. 23740208, 25003345 (M.Y.), 21111006, 23540327 26105520 (K.K.), 
22244030, 26247042 (K.I. and K.K.), 24103006, 24000004 (K.I.), and 
23104006 (M.M.N.), and supported by the Center for the Promotion of 
Integrated Science (CPIS) of Sokendai (1HB5804100) (K.I. and K.K.).


\end{document}